\documentclass[sigconf]{acmart}
\AtBeginDocument{%
  }

\copyrightyear{2025}
\acmYear{2025}
\setcopyright{acmlicensed}\acmConference[SC Workshops \'25]{Workshops of the International Conference for High Performance Computing, Networking, Storage and Analysis}{November 16--21, 2025}{St Louis, MO, USA}
\acmBooktitle{Workshops of the International Conference for High Performance Computing, Networking, Storage and Analysis (SC Workshops \'25), November 16--21, 2025, St Louis, MO, USA}
\acmDOI{10.1145/3731599.3767549}
\acmISBN{979-8-4007-1871-7/2025/11}

\acmSubmissionID{ws\_sfwm109}

\usepackage[english]{babel}
\usepackage[utf8]{inputenc}
\usepackage{lipsum}
\usepackage{graphicx}
\usepackage{listings}


\begin{document}

\title{Towards a user-centric HPC-QC environment}

\author{Aleksander Wennersteen}
\authornote{Corresponding authors}
\email{aleksander.wennersteen@pasqal.com}
\orcid{0009-0006-5486-0980}
\affiliation{
  \institution{Pasqal}
  \city{Palaiseau}
  \country{France}
}
\author{Matthieu Moreau}
\email{matthieu.moreau@pasqal.com}
\orcid{0009-0001-7886-3426}
\affiliation{
  \institution{Pasqal}
  \city{Palaiseau}
  \country{France}
}
\author{Aurelien Nober}
\email{aurelien.nober@pasqal.com}
\orcid{0009-0006-8442-1401}
\affiliation{
  \institution{Pasqal}
  \city{Palaiseau}
  \country{France}
}
\author{Mourad Beji}
\email{mourad.beji@pasqal.com}
\orcid{0009-0008-7998-6357}
\authornotemark[1]
\affiliation{
  \institution{Pasqal}
  \city{Palaiseau}
  \country{France}
}

\date{\today}

\begin{abstract}
    Robust execution environments are important for addressing key challenges in quantum computing, such as application development, portability, and reproducibility, and help unlock the development of modular quantum programs, driving forward hybrid quantum workflows.
    In this work, we show progress towards a basic, but portable, runtime environment for developing and executing hybrid quantum-classical programs running in High Performance Computing (HPC) environments enhanced with Quantum Processing Units (QPUs).
    The middleware includes a second layer of scheduling after the main HPC resource manager in order to improve the utilization of the QPU, and extra functionality for observability, monitoring, and admin access.
    This approach enables managing multiple programming Software Development Kits (SDKs) as first-class citizens in the environment by building on a recently proposed vendor-neutral Quantum Resource Management Interface (QRMI).
    Lastly, we discuss and show a solution for the monitoring and observability stack, completing our description of the hybrid system architecture.
\end{abstract}


\maketitle

\section{Introduction}

Quantum computing appears to be a promising next step for accelerating computation and unlocking new opportunities for select workloads.
It is expected to play a role in the future of supercomputing~\cite{future_of_supercomputing}.
Quantum Processing Units (QPUs) do not operate in isolation.
They require classical computing for essential capabilities such as control, calibration, orchestration, and monitoring.
Moreover, quantum computing is unlikely to provide uniform speed-up across an entire application, thus, requiring integration into larger workflows dominated by classical computing.

Integrating QPUs into HPC environments is therefore essential, not only to enable quantum computing in isolation, but also to make it possible to embed quantum subroutines into state-of-the-art computational science workloads that solve real-world problems~\cite{ALEXEEV2024666}.
In this work we have focused on integrating an analog neutral atom Quantum Processing Unit (QPU) by the vendor Pasqal~\cite{pasqal_whitepaper}. Such QPUs are currently installed in the HPC centers CEA/GENCI in France, JSC in Germany, with another expected in 2026 at CINECA, Italy.

QPU integration must extend beyond compute orchestration to encompass the full system environment.
This includes the monitoring and observability stack of the HPC infrastructure—enabling system administrators to track current and historical device status using familiar tools.
It must also provide access to relevant telemetry for end-users, and support staff, as effective debugging and performance analysis would otherwise be infeasible.

Finally, modern developers have expectations of testing, debugging and development environments.
However, QPUs are currently a much too scarce resource to be able to justify using the real QPUs for such purposes.
This means that developers will have to create programs in one environment, and execute it in another.
This is error prone, especially in quantum programs where the desired backend and its properties are typically defined in source code.
The need for the development of development environments and tooling has already been identified in the literature~\cite{qse_roadmap_challenges}. 

We introduce a new HPC-QC integration stack.
Our main contribution is the design of a new hybrid HPC-QC runtime environment on top of the QRMI interface recently introduced~\cite{qrmi_paper}, and the propagation of this approach through the stack, making both cloud-based and on-prem QPUs that are integrated into HPC environments first class citizens of the stack.
This runtime environment, enabled by a intermediate service daemon, also enables the inclusion of interoperable APIs, allowing third-party components such as optimal control frameworks, calibration tools, or error-mitigation services, to be integrated at the calibration and runtime layers to extend the overall system functionality.
Building on this we outline how we should be able to emulate the final execution environment during development, testing, and trial runs.

We approach this technically by extending the framework of the QRMI interface to locally running emulators, and introduce a new REST API running as a daemon process on the quantum access node.

This remainder of this paper is organized as follows: Section~\ref{sec:background} provides the motivation and prior art on development and execution environments for quantum computers, HPC-QC integration, runtime environments and middleware, and fine-grained scheduling techniques to improve QPU utilization.
Section~\ref{sec:architecture} presents the general architecture of the proposed solution.
Section~\ref{sec:observability} discusses the observability and monitoring dimension.
Finally, Sections~\ref{sec:discussion} and~\ref{sec:conclusion} conclude with a discussion, future work, and summary of the results presented.

\section{Background}
\label{sec:background}

\subsection{Development and execution environments for quantum computers}
\begin{figure*}
    \centering
    \includegraphics[width=\linewidth]{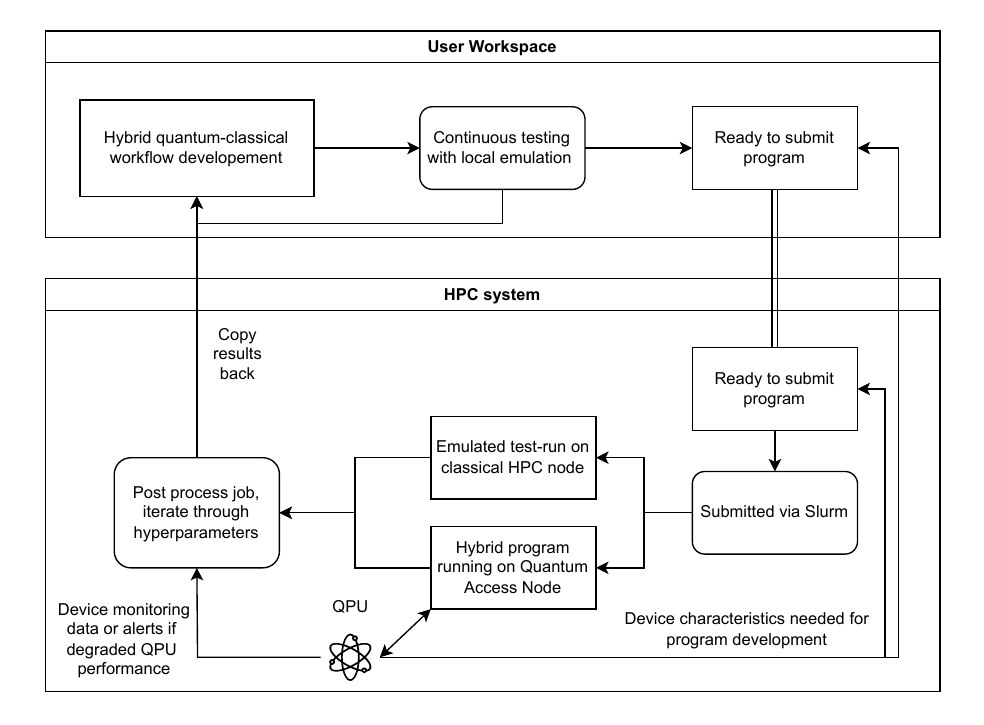}
    \caption{Overview of the HPC–QC development workflow. Programs typically move from local development, where continuous testing with emulation is possible, to execution on HPC systems. Within the HPC environment, jobs may run either on classical emulators or on the QPU. At each stage, users must update device targets and fetch current hardware characteristics, which complicates maintaining a single, unchanged program across environments. }
    \label{fig:hybrid-workflow-overview}
\end{figure*}

As shown in Figure~\ref{fig:hybrid-workflow-overview}, to run a hybrid quantum-classical program there are many stages the user must go through.
These are all different development or execution environments the user must set up and keep compatible for the length of the development lifecycle.
In the HPC setting, traditionally this has been mitigated by encouraging development on the cluster.
However, it is often difficult to set-up a modern development environment on in such a setting, as well as using cloud resources for early testing.
The need for improved tooling and development environments in quantum computing has been acknowledged in recent literature~\cite{qse_roadmap_challenges}, to help mitigate the negative effect of this cycle among other reasons.
In particular, challenges related to environment setup, portability, and testing have been highlighted in empirical studies and surveys~\cite{Hevia2024psy}.

When changing devices, from emulators to QPUs, existing solutions require untested code paths to handle different execution backends via source code changes.
Even without changing devices, a key difficulty in hybrid quantum-classical program is the lack of portability of quantum programs across different environments.
Current quantum development workflows often assume tightly coupled development and execution stages, which makes it difficult to iterate or migrate programs~\cite{Hevia2024psy}. 
Separating development layers from execution, as opposed to decoupling job submission from resource usage, can mitigate this challenge.
For instance, such separation is partially addressed by mechanisms at the scheduler level that allow separating the quantum resource definition from the source code of the program~\cite{qrmi_paper}.

That said, quantum systems introduce new constraints. The compilation and execution stages are often coupled with device calibration status and runtime environment specifics.
Some proposals emphasize the need for separation between compilation and runtime~\cite{towards_hybrid_toolchain}, but this is typically considered from a static compilation viewpoint.
In practice, quantum programs are developed in Python and rely on interpreted environments and runtime code generation. Our perspective emphasizes usability for current workloads, particularly the developer's experience as they transition from prototyping to large-scale execution across heterogeneous systems.

In terms of portability current, pre-fault-tolerant, quantum computers have an additional temporal dimension to consider compared to traditional computers: 
quantum processors are subject to calibration drift over time, which can lead to discrepancies between the environment in which a program is developed or tested and the one in which it is executed.
This issue is particularly acute in programmable quantum simulators, where device parameters significantly affect program semantics.
Ensuring program validity at the point of execution thus becomes a key requirement.

While foundational research has begun to define testing frameworks and techniques~\cite{quantum_software_testing_shaukat, debugging_quantum_programs}, practical tools have only recently started to emerge in user-facing frameworks~\cite{rovara2024frameworkdebuggingquantumprograms}.
Although these developments mark important progress in testing of the pure quantum portion, testing hybrid programs at scale remains a problem.
One can perform end-to-end using emulators as will be discussed in section \ref{sec:portability}.

\subsection{HPC-QC Integration}

Quantum processors are not standalone devices; they rely on a tightly integrated classical infrastructure for control, calibration, and the execution of hybrid algorithms.
Since quantum computers are not expected to outperform classical systems across all problem domains, their integration into high-performance computing (HPC) environments is a natural strategy to realize practical quantum advantage~\cite{ALEXEEV2024666}.

The integration of quantum and HPC resources appears, at its core, a software problem~\cite{software_challenge}. While progress in quantum hardware is essential, we do not foresee a requirement for specialized physical-hardware interfaces. All current QPUs require classical control orchestrating microwave or laser pulses which is the lowest possible integration points with HPC orchestration layers. Given the difficulty of building large-scale quantum computers and error correction, the integration point is likely to be even higher in the stack.

\subsubsection{Integration tightness}
\label{sec:integration-tightness}
The integration of quantum and classical compute resources is not solely motivated by workflow orchestration. It is also needed, for example, for quantum error correction decoding. However, it remains an open question to decide the extent to which the required classical compute should reside within the HPC system versus be provided via dedicated infrastructure attached to the quantum device itself.

A useful distinction in the integration landscape is between loose and tight coupling models~\cite{HPC_with_QPUs}. Loose coupling typically involves cloud-hosted QPUs accessed asynchronously, whereas tight coupling refers to on-premises or even on-chip integration enabling low-latency classical-quantum interaction. The tight coupling model underpins the viewpoint of "quantum accelerators"~\cite{quantum_accelerators_for_HPC,QCs_for_HPC}, where QPUs function similarly to GPUs and are addressed through familiar offloading APIs.

We note that the time scales of operation on different devices differ by several orders of magnitude.
For current neutral-atom devices, the shot rate is on the order of 1 Hz, with roadmaps projecting increases to around 100 Hz in the coming years\footnote{See e.g. Pasqal roadmap webinar \url{https://www.pasqal.com/events/pasqal-quantum-roadmap-update-webinar/}}.
Due to these time scales, we do not consider tight integration, which would require low-latency or high-bandwidth connections, to be a practical concern in the near term, as no such bottlenecks have been observed in our experience.

\subsection{Runtime environment and middleware}%
\label{sec:runtime-middleware}

Prior work in this area has primarily focused on tightly coupling a single software stack to hardware backends~\cite{integration_of_quantum_accelerators, MCCASKEY2018245, mqss}, or
on resource utilisation optimization rather than enhancing the user experience~\cite{viviani2025assessingelephantroomscheduling}.
To address this, we will advocate for a coherent multi-SDK execution environment, structured around a unifying runtime layer that interfaces with the scheduling middleware.

We distinguish between:
\begin{itemize}
    \item \emph{runtime}---what executes a single quantum or hybrid program instance on a specific device, and provides the portability functionality
    \item \emph{middleware}--- what orchestrates many such program instances across heterogeneous quantum-classical resources, manages workflows, and handles multiple concurrent users.
\end{itemize}
In practice, the two layers often blur as a coherent runtime environment, but treating them separately clarifies the design space discussed below.
In the case of QPUs integrated with HPC systems, the runtime and middleware can further interact with the HPC environment.
Using more comprehensive middlewares, collaborating with the HPC environment, we also envision being able to facilitate more clever scheduling with concurrent user jobs. 

Many system-level frameworks such as XACC~\cite{MCCASKEY2018245} or the Munich Quantum Software Stack~\cite{mqss} or CUDA-Q adopt a compiler-centric
view: kernels written in various front-end languages are lowered to an intermediate representation and then dispatched
to a backend (QPU).

Although the above efforts demonstrate the feasibility of site-wide quantum middleware,
none of them targets the multi-SDK use case highlighted in
Section~\ref{sec:multi-sdk}.
In practice, most quantum developers today use interpreted languages like Python, which shifts the development focus away from the potential optimizations of compiler-led approaches.

\subsubsection{Coherent Multi-SDK Environments} \label{sec:multi-sdk}

The quantum software ecosystem remains fragmented, with limited standardization across programming interfaces. A single quantum processing unit (QPU) may be programmable through multiple SDKs, each with distinct abstractions, runtime behaviors, and levels of hardware integration.
For instance, QPUs by Pasqal can currently be accessed via Pulser~\cite{pulser}, Qiskit\footnote{\url{https://github.com/qiskit-community/qiskit-pasqal-provider}}, CUDA-Q, and Qaptiva/QLM.

While this diversity provides flexibility and accessibility to different developer communities, it also introduces significant challenges. From the perspective of users, hosting sites, and hardware vendors, maintaining and supporting a growing number of software stacks increases operational complexity. Inconsistent runtime behavior, dependency management, and integration effort can hinder reproducibility and limit the portability of quantum applications.

\subsection{Multi-level scheduling for improved resource utilization}
\label{sec:scheduling}

\begin{table*}[t] 
  \centering
  \begin{tabular}{@{} l c c l @{}}
    \toprule
    \textbf{Pattern} & \textbf{Quantum load} & \textbf{Classical load} & \textbf{Scheduler hint} \\
    \midrule
    \textbf{A) High-QC / Low-CC} & Dominant & Minor pre/post processing & Sequential QPU queue \\
    \textbf{B) Low-QC / High-CC} & Sparse   & Heavy                    & Interleave jobs to kill QPU idle time \\
    \textbf{C) Balanced QC-CC}   & Comparable & Comparable              & Fine-grained orchestration \\
    \bottomrule
  \end{tabular}
  \caption{Taxonomy of hybrid quantum–classical workload patterns and associated scheduling strategies. The table highlights three common workload types: QPU-dominant, CPU/GPU-dominant, and balanced. It suggests scheduler hints that can reduce idle time and improve overall resource utilization.}
  \label{table:taxonomy}
\end{table*}

Recent work shows that substantial improvements to resource utilization is possible by allowing the application to dynamically grow or shrink at run time, so-called malleable jobs~\cite{viviani2025assessingelephantroomscheduling}.
This requires that the resource manager exposes suitable APIs and that the applications co-operate. And despite significant research over several years such an approach has yet to become widely available in classical HPC contexts~\cite{malleability_in_hpc}.

Other recent work, upon which we also base our taxonomy in Table \ref{table:taxonomy}, provides a software stack perspective and proposes that any HPC-QC scheduler must consider workload patterns and support multiple types of allocation~\cite{bridging_paradigms}.
In summary, the literature converges on the view that flexible, pattern-aware, and ideally malleable scheduling is essential to unlock the real utilization of scarce QPUs while safeguarding classical efficiency.

As argued in Ref.~\cite{viviani2025assessingelephantroomscheduling}, QPUs are currently a very scarce resource in HPC systems. Nevertheless, it is important to avoid wasting large amounts of classical HPC resources.
As the machines grow in size the post-processing of bitstrings become more resource intensive. For example in the recently introduced Sample-based Quantum Diagonalization approach (SQD)~\cite{sqd}, where the post-processing was parallelized up 6400 nodes on Fugaku.

As we can see in Table \ref{table:taxonomy}, the intensiveness of the quantum and classical portions of quantum workloads vary. Given the scarcity of quantum resources it is often desirable to maximize the QPU utilization. However, the wall-clock runtime of the QPU runs can often be large. This causes under-utilization of classical resources. An intermediate scheduler allowing for more customization in the scheduler can unlock solutions to this, such as malleability~\cite{viviani2025assessingelephantroomscheduling}.
On top of this, development jobs, testing etc, can be interesting to enable as a low-priority job queues.

\subsection{Observability and low-level controls} \label{sec:background-observability}

Although still experimental, quantum computing is growing at a fast pace and many different actors are already able to produce various components of the hardware and software stack. It is key for HPC centers to be able to make these various components work together, and the requirements in terms of interoperability are growing. A few examples are devices for calibrating quantum operations using techniques such as optimal control, or systems meant to learn and correct noise. 

These components often require access to APIs at levels much lower than what users need to program devices.
Exposing low-level controls of QPUs is not always safe, however, due to the complexity of the hardware and these controls are often provider-specific.
Exposing a subset of these low-level APIs and having the ability to implement increased safeguards should be performed at the daemon level.
This indirection provides a natural point to define interoperable APIs and integrate third-party components, enhancing QPU calibration, performance, and runtime features.

Often not discussed are the challenges in monitoring QC systems for HPC centers.
Unlike classical nodes, which exhibit predictable and mature operational behavior, quantum systems remain
susceptible to fluctuations in calibration parameters, environmental noise, and device stability over time.
These variations, whether in qubit coherence times, gate fidelities, or laser alignment, can directly affect job accuracy and runtime efficiency.
Recently, a comprehensive framework was introduced to monitor QPUs installed in data centers~\cite{telemetry_for_quantum_systems_in_HPC_centers}.

For HPC operations teams, this introduces a new observability dimension.
They must be able to track QPU health in real time, detect degradation trends and schedule maintenance as part of their day-to-day work.
For end-users, transparent reporting—such as per-job metadata on qubit performance can assist in interpreting noisy results and guide adaptive workflows. 

\section{Architecture} \label{sec:architecture}

\begin{figure*}
    \centering
    \includegraphics[width=\linewidth]{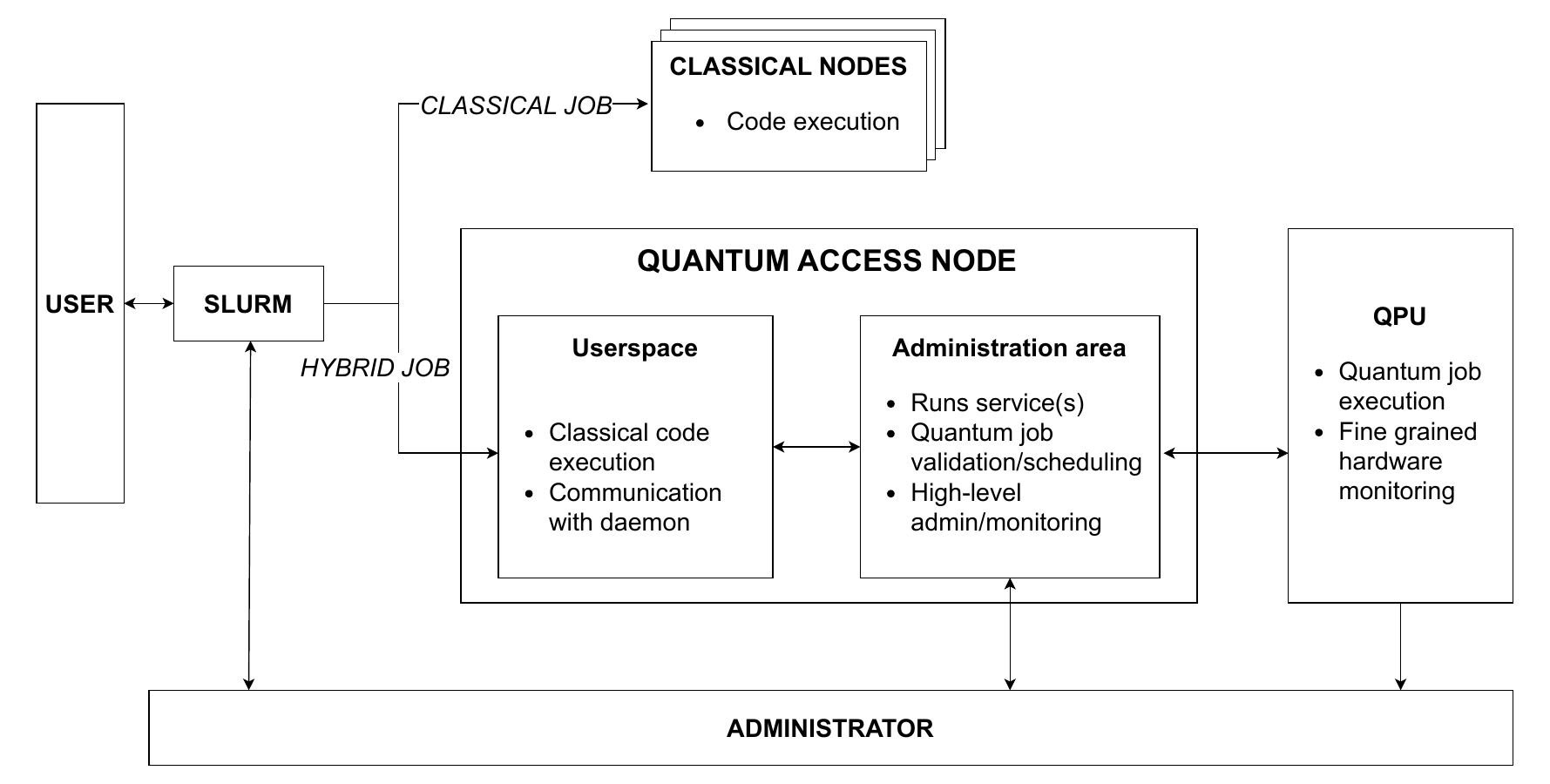}
    \caption{Architecture of the overall HPC-QC integration approach. A dedicated Quantum Access Node hosts the QPU connection and runs a lightweight middleware daemon on reserved resources. This daemon mediates user interactions across multiple SDKs, manages validation, prioritization, and scheduling of quantum tasks, and integrates with the HPC batch scheduler. Classical nodes continue to execute CPU/GPU workloads or the classical parts of hybrid jobs. Administrative connections are available at all layers, enabling the QPU to be managed as a peripheral resource within the HPC system. The design is modular and flexible, allowing adaptation to the policies and requirements of individual hosting sites.}
    \label{fig:solution-overview}
\end{figure*}

Building on these insights we advocate a shared, multi-user, runtime scheduling layer that sits between the batch scheduler (e.g. Slurm) and the QPU,
enabling multi-level scheduling and facilitating more advanced techniques.
Such an approach enables both further practical research and development into malleable applications, yet provides initial benefits to users today. We show a schematic of the solution and how it relates to the HPC infrastructure in Figure~\ref{fig:solution-overview}.

Concretely, we use the QRMI as our primary unifying runtime library interface. Our work extends the QRMI framework from providing connectivity and Slurm scheduling, with a second level of scheduler capability with the middleware daemon.

We aim to not be HPC-exclusive, as the final solution needs to work also outside HPC environments.
In our experience this move between environments is very common and significantly complicates the user experience and reinforces the survey findings in~\cite{Hevia2024psy}, which highlight the difficulty of preparing consistent execution environments. 
Providing a runtime environment abstraction that behaves similarly on HPC systems and locally on the developers machine is expected to significantly reduce common bugs observed.

\subsection{Runtime environment}
\label{sec:runtime}
On top of the QRMI-based Slurm plugin, which provides the HPC-facing runtime API, 
we introduce a dedicated runtime environment tailored for hybrid quantum–classical applications. 
This layer is responsible for executing jobs across CPUs, GPUs, and QPUs while abstracting away 
low-level details such as hardware availability, queuing mechanisms, and backend selection. 
For developers, the runtime provides a consistent interface that supports transparent switching 
between high-performance emulators and physical QPUs. 
This ensures that the same workflow can be used for development, testing, and production, 
significantly improving portability and reducing friction when debugging or validating hybrid programs.

While the runtime ensures that a single application can be executed consistently in different environments, effective integration in multi-user HPC environments 
requires an additional coordination layer. 
This role is fulfilled by the middleware, which manages concurrent sessions, 
prioritization, and system-wide policies. We will discuss this in section \ref{sec:middlware}.

First, we will see how emulation enables early-stage development and validation, then we describe how the same interface exposes real QPU and cloud resources, ensuring portability across development and production environments.

\subsection{Seamless portability from development to production environments}
\label{sec:portability}

Leveraging the functionality provided by QRMI, we expose as devices to the scheduler and enable switching via \texttt{--qpu=<resource>} the following devices:
\begin{enumerate}
  \item On-premises QPU connection
  \item Cloud-based QPU resources
  \item Cloud based emulator resources.
\end{enumerate} 

Additionally, we implement as a \texttt{QRMIBackend} the emulator suite from Ref.~\cite{emulators}.
The user-exposed backend module will default to using the tensor network backend, if installed.


During development or when QPU access is constrained, users can target a high-performance emulator
that replicates the expected quantum results.
At runtime, a single configuration change with the \texttt{--qpu} option instead sends the job to physical hardware.
This interchangeability simplifies reproducibility and ensures
that workflows remain portable between development and production environments.

By defaulting to execution on our open-source emulators the user is able, assuming sufficiently small quantum programs, to run their program locally on their laptop, then move it to the HPC cluster and test on large tensor network emulators before running the final job on the QPU, using the Slurm plugin to move between jobs.

To the best of our knowledge there are not any targeted efforts on facilitating the testing or portability of quantum applications or workflows.
This work partially addresses this by enabling testing with tensor network emulators for the QPU considered provided by the vendor~\cite{emulators}.
By restricting the bond dimension, tensor network emulators can execute programs on almost arbitrarily large\footnote{In the limit of not allowing any entanglement to be represented, i.e. restricting the quantum state to a product state, 2 complex numbers must be stored per qubit. The evolution of such a restricted system is not interesting, but it can be used for mocking the QPU in end-to-end tests.} QPU emulators.
Although the result will not be accurate,
this allows for validating the hybrid program against the current device state by fetching its up-to-date specifications.

Together, this provides a simple realization of the vision in Figure~\ref{fig:hybrid-workflow-overview}, allowing the end-user to develop and test their workflow without changing the program.

\subsection{Middleware service}
\label{sec:middlware}
By introducing a simple service exposed as a RESTful API, limited to managing the currently running jobs and sessions of the QPU, we insert an abstraction layer between user sessions and the QPU task queue.
This enables a second level of scheduling logic that allows multiple users to share the QPU, while also providing a single configuration point for concurrent sessions.
An additional benefit is the ability to customize features,
extending or modifying scheduling policies, which would otherwise be locked within the vendor's proprietary software stack.
We discuss such advanced scheduling ideas in section \ref{sec:scheduling}.

Once the operation of quantum computers becomes more standardized, the functionality of this service can be ported to run directly on the QPU controller rather than as a standalone process.

For the hosting site, the daemon-based model localises device access logic and reduces the
maintenance burden that currently increases with the number of supported SDKs.
For users, it preserves the development experience while enabling transparent migration from development runs on simulators
to production runs on physical QPUs.

We construct these services building on our existing cloud infrastructure~\cite{bidzhiev2023cloudondemandemulationquantum}, ensuring interoperability between the on-prem QPUs and cloud-based resources.

\paragraph{User sessions and job priorities}
As the user part of the runtime environment connects to the middleware, a unique session is created, and a session token is returned.
The daemon receives jobs from one or more sources. A priority queue is implemented, for example we can envision several classes of user jobs:

\begin{enumerate}
    \item production jobs (top priority)
    \item test runs / scalability tests (medium priority)
    \item development runs (low priority)
\end{enumerate}

The production job should always be able to pre-empt running jobs of lower priority automatically, although the initial implementation such sharing is implemented by having non-production jobs configured with a low number of shots and without batched submission.
This ensures that the waiting time for production jobs will be low, and the scheduling algorithm will always schedule the highest priority job.
Although not part of this work, the system could be extended to also accept jobs via a cloud interface, similar to how it is handled in the JHPC-Quantum\footnote{\url{https://jhpc-quantum.org/en/about/}} project. 

A first level of queueing is provided by Slurm.
The different job priorities also correspond to Slurm partitions, which should be assigned different priorities.
A second layer of scheduling, and multi-user-management for optimal resource utilization, is provided by the middleware daemon.The daemon retrieves the job's priority from Slurm.

\subsection{Installation and configuration}

The daemon service must be installed in a way not accessible to users except for with the exposed REST API, and with dedicated resources.
This can be done at minimum using Linux cgroups, but where possible other solutions like containers or virtual machines should be considered.

Since QRMI is configured through environment variables, it is natural to rely on configuration files and environment settings. These can be defined at different levels: locally by developers, within an integrated development environment, or automatically by the HPC scheduler.
The hosting site can modify both the number of queues and their relative priority; similar to the Slurm \texttt{slurm.conf} configuration file.
We discuss more advanced solutions to this in Section \ref{sec:discussion}.
Additionally, quality assurance jobs checking the QPU is typically scheduled periodically by both the hosting site and the QPU itself for internal checks using the QPU scheduler directly.

Using the QRMI we implement this interface also for numerical emulators of the QPU. QRMI already supports Qiskit and Pulser backends, and Slurm Spank plugins, with GRES plugins being actively worked on. This must be configured in the configuration file as well.

\subsection{Advanced scheduling}
Scheduling should be done building on top of the HPC system's scheduler. Focusing on Slurm one could use either licenses or GRES resources to assign partial QPU resources.
Without requiring changes to Slurm, we could in both cases assign 10 licenses/GRES units, corresponding to timeshares of the QPU in increments of 10 percentage points.
The QRMI can be modified to communicate this proportion of QPU share to the scheduler on the middleware service daemon.

Interesting future additions include intelligent scheduling such as pattern-aware scheduling and interleaving based on whether a job is QC-dominant, CC-dominant, or balanced.
This is crucial for avoiding idle QPU time, which is particularly costly given the rarity and value of quantum resources.
We could for example enable adding \texttt{--hint=qc-balanced}, and others as listed in Table \ref{table:taxonomy}, or even more detailed information such as the expected
time running on the QC hardware to enable planning interleaving other jobs more effectively.
Most likely this is best done inside the QPU, or using solution such as the daemon service presented herein.
In that case both the programming model and resource manager could be imagined to give such hints.

\subsection{Observability and low-level controls} \label{sec:observability}

In this work we build a native observability stack, exposing QPU state through standard
telemetry tools such as Prometheus, with plans to integrate dashboards via Grafana, all built on the InfluxDB time series database.
Using such standard tools makes it easy to integrate the QPU metrics into existing observability stacks at the data center.

This information will also be exposed to the user where appropriate. Which will aid in understanding the job behavior.

In this case, just as with the scheduling daemon, by exposing this to a stand-alone daemon that interacts with the QPU we enable a much larger level of control for the hosting site than could be possible by a monolithic approach where everything runs on the QPU.

\section{Discussion}
\label{sec:discussion}

The architecture proposed in this paper is intentionally minimal and modular.
Rather than building a monolithic software stack, we aim to isolate integration pain points—multi-SDK fragmentation, mismatched scheduling granularity, and developer workflow friction—and address them through composable components that fit into existing HPC practices.

\paragraph{Modularity over vertical integration.}
Whereas currently available software often assumes vertical integration between SDK, runtime, and even control hardware, our approach allows sites and users to mix and match components.
A key enabler of this modularity is the lightweight RESTful API service, which decouples user interactions from the vendor stack and exposes a clean interface for extending functionality such as scheduling policies or observability hooks.
This is essential in the pre-standardization phase of quantum software, where multiple frameworks coexist and evolve independently.
This modular approach, with the right integrations, will also allow third-party developers to configure or extend it.

\paragraph{User and developer experience as the top priority.}
Improving the developer and user experience, especially with respect to improved ability to test and develop hybrid workflows across multiple execution environments was the top priority in this work.
We solved this by introducing a simple runtime environment and middleware built on top of an open source Quantum Resource Management Interface that integrates with Slurm.
We discussed extending this to multiple execution environments, including outside of the HPC context.

\paragraph{Scheduler collaboration, not replacement.}
We deliberately designed our runtime to complement existing batch scheduling infrastructure (e.g., Slurm), not replace it or make major modifications.
This ensures that our system can be deployed with minimal friction and supports familiar workflows for both users and operators.
However, richer two-way communication between the scheduler and the runtime would enable deeper coordination between the scheduling of the quantum partition and the rest of the HPC resources.
For example, exposing lower-level job state transitions or predicted QPU durations could allow more aggressive backfilling and fairer resource sharing.

\paragraph{Emulation and testability.}
Providing transparent access to software emulators through the same runtime interface as real QPUs is a pragmatic solution to the resource scarcity faced by developers.
However, the current emulator modes are coarse-grained and best suited to functional validation, not performance evaluation.
Future efforts could enrich the emulator interface with profiling, fault injection, or simulated QPU timing to enable more realistic development.

\paragraph{Limitations and future directions}
While our current implementation demonstrates multi-SDK support and emulator fallback, it targets a single QPU vendor.
Extending to a true multi-vendor setup will require careful abstraction of backend capabilities and metadata.
Similarly, we currently assume HPC environments and Slurm integration.
Adapting this to cloud-native execution models would be important for broader adoption, the current implementation is limited to accessing cloud QPUs from HPC environments.
The Pasqal QPU operates in the analog regime, with a roadmap towards digital, fault-tolerant, devices. The current implementation is restricted to production devices. It can be extended to digital devices once these become generally available. The HPC-QC integration model is also assumed to be loose for reasons discussed in section \ref{sec:integration-tightness}.

Future work should explore tighter integration with batch schedulers, pattern-aware scheduling policies, and richer co-scheduling strategies for hybrid jobs. It should integrate better with the existing HPC environments, and through collaboration with partners better support innovative solutions for scheduling, for example via workflow engine integrations or malleable jobs.
When extending support to digital, and especially fault-tolerant, devices, the architecture should, for example, address the compilation process.
For an example of such features, see \cite{mqss}.
To preserve modularity, this functionality should primarily be developed upstream in shared projects such as QRMI. 
Finally, while emulator fallback improves testability, adding explicit debugging, profiling, and reproducibility tools would significantly enhance the developer experience across heterogeneous environments.

On the observability side, extending beyond basic telemetry toward per-job metadata and automated drift detection would further improve system reliability. The observability stack should be better integrated with the software stack.

\section{Conclusion}
\label{sec:conclusion}

Our proposed architecture introduces a lightweight runtime environment built on top of the vendor-neutral Quantum Resource Management Interface (QRMI) connected with a middleware that will handle concurrent and prioritized multi-user scheduling, admin operations and monitoring. The daemon-based middleware, running as a standalone REST service on quantum access nodes, transparently unifies multiple quantum SDKs using a common runtime interface based on the QRMI without requiring significant architectural changes to user workflows or vendor software. It integrates directly with the HPC batch scheduler (e.g., Slurm), leveraging existing scheduler configurations such as partitions and GRES resources to handle prioritization and resource management. Additionally, the architecture seamlessly supports emulator-to-QPU transitions, simplifying the development-to-production workflow. The strengths of our design include modularity, SDK-neutrality, compatibility with established HPC practices, and enhanced observability through standard telemetry tools like Prometheus and Grafana.

The design is compatible with local execution during development as well as with both cloud-based and on-premises QPUs. It supports seamless transitions between local development, HPC emulation, and production QPU execution environments, which we believe will significantly improve developer productivity and user experience.

While our current implementation targets a single vendor and Slurm-based systems, the architectural principles are general. Future work will incorporate more scheduling metadata, and explore richer co-scheduling models with tighter system-scheduler integration. By integrating with existing job schedulers such as Slurm and surfacing quantum-aware scheduler hints, one can bridge the gap between current quantum development tools and production HPC workflows. This would enable more efficient resource utilization, improve developer experience, and reduce integration burden for hosting sites.

\section*{Code Availability}
A reference implementation is planned for release as open-source software on \url{https://github.com/pasqal-io}.
Where possible we aim to upstream all relevant code.

\begin{acks}
The authors thank our partners at CEA, GENCI, Jülich Supercomputing Center, and CINECA for constructive discussions on the on-prem integration of Pasqal's QPUs into their respective centers.
\end{acks}
\bibliographystyle{ACM-Reference-Format}
\bibliography{ws_sfwm109}
\end{document}